\documentstyle[epsfig,11pt]{article}

\def\beq{\begin{equation}}
\def\eeq{\end{equation}}
\def\noi{\noindent}
\def\lra{\leftrightarrow}
\def\pd{\partial}

\title{\bf BKP states in the inclusive gluon production}
\author{M.Braun \\
Department of high-energy physics,\\
University of S. Petersburg, 198904 S. Petersburg, Russia }
 \date{March 2010}

\begin{document}
\maketitle
\medskip
\noi{\bf Abstract}
Inclusive cross-section for gluon production is calculated by the
dispersion technique in the NLO in the perturbative QCD with a large
number of colours. The found cross-section coincides with  the one
derived in the dipole picture. No trace of the BKP states is found.

\section{Introduction}
In the perturbative QCD the inclusive gluon production is one of the most
important observables, which can be directly compared to the experimental data.
Applied to interaction with heavy nuclei it was first studied in
~\cite{bra1} using the standard AGK rules and neglecting emission from the
triple pomeron vertex. Later Yu. Kovchegov and K.Tuchin derived the single inclusive
cross-section from the dipole picture ~\cite{kovch}. Their expression contained an
extra term as compared to ~\cite{bra1} which was shown to correspond to emission from
the triple-pomeron vertex ~\cite{bra2}. However recently J.Bartels, M.Salvadore and G.-P.
Vacca performed a new derivation ~\cite{BSV} based on the dispersion approach,
which studies discontinuities
of amplitudes in various $s$-channels, the BFKL-Bartels (BB) technique
 ~\cite{BFKL},~\cite{bar} .
Their results are different from the previous ones
in ~\cite{kovch}. They contain new terms of a complicated structure, which seem to involve
the so-called BKP states, higher pomerons composed of 3  and 4   reggeized
gluons.
Presence of such states  will inevitably make calculations of the inclusive cross-section
much more difficult (if possible at all), since their wave functions  depend
on many variables and are unknown at present. So the problem to understand the origin
of the difference in the two derivations is quite important.

It may be that after all the initial physical picture in the two approaches is
different. In the BB picture one is studying discontinuities of the standard Feynman
diagrams in the Regge kinematics. In the Kovchegov-Tuchin (KT) picture  gluon emission
occurs due to eikonal scattering of fast quarks and gluons passing through the
nucleus with instantaneous interactions with its components. So the first task in the
study of the difference between the two results is to compare the initial
expresions,
which combine into the final inclusive cross-section. If they are different then in
fact the two pictures are physically different and the difference in the
resulting cross-sections is natural.

Actually the difference between the two pictures starts from the
next-to-leading order in $\alpha_s\ln s$ (NLO) and originates from the contribution
of of one extra gluon softer than the observed gluon.
One can prove that in this order the
tree diagrams which describe single and double gluon emission identically
coincide in these two pictures.  The inclusive cross-section also includes
diagrams with loops formed by the gluon softer than the observed one.
In our previous paper
~\cite{braun09} we were able to demonstrate that for single gluon emission and
single interaction with the target contributions from such
loops  also identically coincide in the BFKL and the Kovchegov-Tuchin
techniques. In the BFKL approach the mentioned loop generates
the Regge trajectory of the exchanged gluon and thus is responsible for
gluon reggeization. Our results showed that gluon reggeization is in fact
realized by the loop contributions to the emission during eikonal scattering.
More complicated loop diagrams
with two or three interactions with the target also appear
in the inclusive cross-section in the NLO.
They are numerous and their full study is still far from being accomplished.
However comparison of a relatively small subset of such diagrams,
distinguished by their special dependence on the
observed gluon momentum and polarization shows
that their contribution again fully coincides in the two formalisms.
So it seems almost cerain that the physical pictures described by the
BB and KT formalisms are fully equivalent and one may expect the inclusive
cross-sections derived in these approaches to be completely identical.

The equations derived in ~\cite{BSV} for the inclusive cross-section in
the BB formalism are quite complcated and not at all transparent.
One cannot exclude that in their
solution all contributions which seem to be additional as compared to the
KT result in fact cancel, so that in the end the two cross-section indeed
coincide. In this paper, to study this point, we directly compare the
two cross-sections in the NLO, at which, as mentioned,
the additional contribution starts in ~\cite{BSV}.

Our results show that the cross-sections described by the
equations derived in the BB approach in ~\cite{BSV} and in the KT approach
in ~\cite{kovch} completely coincide. So indeed all the
additional contributions from the BKP intermediate states cancel.

The paper is organized as follows. In the next section we consider
the inclusive cross-section  in the KT approach in the first two
orders. Section 3 is dedicated to a general overview of the
inclusive cross-section in the BB aaproach and the relevant
diagrammatic rules.
Several next sections are devoted to direct calculations of various
contributions to the cross-sections in the two possible target configurations,
the so-called diffractive and double cut ones.
In Sec 10 we collect our results to find the final inclusive cross-section
in the next-to-leading order in the BB approach and compare it with the
KT approach. In Section 11 we draw some  conclusions.
A few details concerning presentation of the KT cross-section in terms
of BB diagrams are given in the Appendix 1. An essential identity for the
BFKL pomeron wave function is derived in the Appendix 2.

\section{Inclusive gluon production in KT approach}
The KT expression for the inclusive gluon production cross-section was
formulated in ~\cite{BSV} in the framework of the dipole picture and
uses the coordinate space formalism. However it can be directly
reformulated in  the BB language in the momentum space
using results obtained in ~\cite{bra2} and ~\cite{brava}. We shall use
the momentum space formulation to facilitate
the final comparison of the KT and BB cross-sections.

\begin{figure}
\hspace*{4 cm}
\epsfig{file=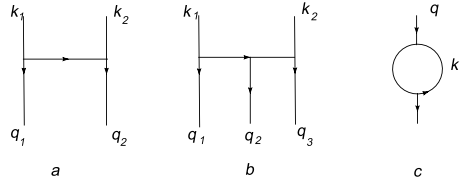,width=6 cm}
\caption{The Lipatov (a) and Bartels (b) kernels.}
\label{fig1}
\end{figure}

In the BB language contributions are described by a set of diagrams
defined in the transverse momentum space and constructed from
effective Lipatov and Bartels kernels, $V$ and $W$ correspondingly,
illustrated in Fig. \ref{fig1}$a,b$ and explicitly given by
\beq
V(q_1,q_2|k_1,k_2)=
\frac{(k_1+k_2)^2}{k_1^2k_2^2}-\frac{q_1^2}{k_1^2(k_1-q_1)^2}
-\frac{q_2^2}{k_2^2(k_2-q_2)^2}
\label{defv}
\eeq
and
\[
W(q_1,q_2,q_3|k_1,k_2)=\]\beq
\frac{(k_1+k_2)^2}{k_1^2k_2^2}-\frac{(q_1+q_2)^2}{k_1^2(k_2-q_3)^2}
-\frac{(q_2+q_3)^2}{k_2^2(k_1-q_1)^2}+\frac{q_2^2}
{(k_1-q_1)^2(k_2-q_3)^2}.
\label{defw}
\eeq
They also  involve the gluon Regge trajectory Fig. \ref{fig1}$c$
defined by the integral
\beq
\omega(q)=-g^2N_c\frac{1}{2}\int\frac{d^2k}{8\pi^3}\frac{q^2}{k^2(q-k)^2}.
\eeq

\begin{figure}
\hspace*{2 cm}
\epsfig{file=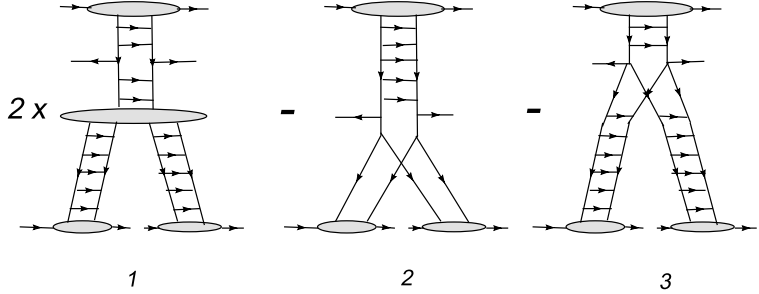,width=10 cm}
\caption{Inclusive gluon production in the KT approach.
Emission from the upper pomeron (1+2) and from the vertex (3).}
\label{fig2}
\end{figure}

\begin{figure}
\hspace*{5 cm}
\epsfig{file=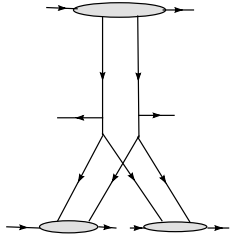,width=3 cm}
\caption{The KT inclusive cross-section in LO}
\label{fig3}
\end{figure}

In this language the total contribution to the inclusive cross-section
splits into two parts: emission from the upper pomeron
and from the first splitting vertex. For scattering
on two centers this generates three sets of diagrams shown in Fig.
\ref{fig2}. The first two diagrams corespond to emission from the
upper pomeron with the following splitting (Fig.\ref{fig2},1) or with
the second term in the expansion of the
Glauber expression for the state to be used as the initial state for
the evolution
in the Balitsky-Kovchegov equation (Fig.\ref{fig2},2).
The third term (Fig. \ref{fig2},$3$) corresponds to emission from the vertex.
In the leading order (LO) the diagram in Fig. \ref{fig2},1 gives no
contribution and the two others reduce to the same diagram shown in
Fig. \ref{fig3}. It has been demonstrated in ~\cite{bra2} that this
LO contribution is exactly reproduced by the sum of all relevant diagrams
in the BB approach.

\begin{figure}
\hspace*{2 cm}
\epsfig{file=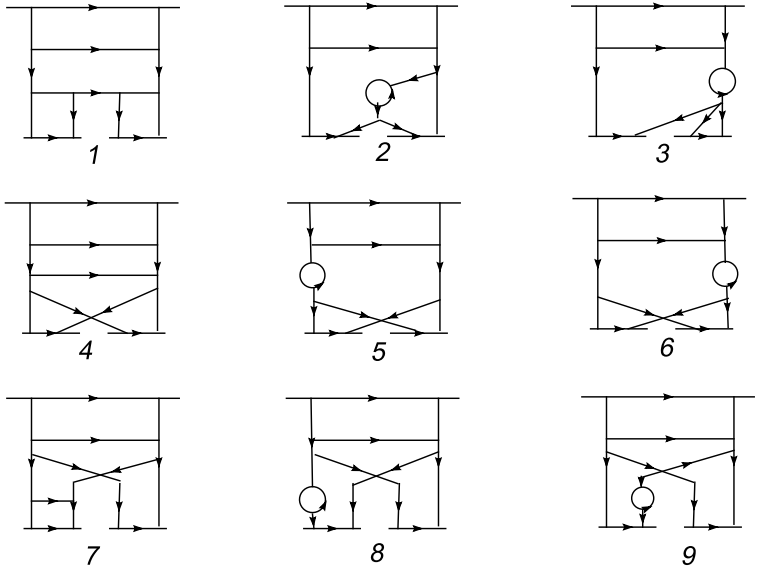,width=10 cm}
\caption{The KT inclusive cross-section in NLO. Vertical lines denote
reggeons. Uppermost and lowermost horizontal lines denote the
colliding quarks. Middle lines denote the observed gluon with momentum $k_1$
(upper) and virtual softer gluon with momentum $k_2$}
\label{fig4}
\end{figure}

In the NLO the KT cross-section is reperesented
by a sum of diagrams shown in Fig. \ref{fig4}.
To simplify, we  consider quarks both as the projectile and the two
targets. It will inevitably introduce some infrared singularities into
the cross-section, which are however irrelevant for our purpose, since the
impact factors for the particpants factorize. To simulate
the colorless quark-antiquark loops we average on the quark colours and
add a similar contribution from the antiquark. To economise on notations
in the end we shall suppress all necessary integrations, obvious powers of the
coupling constant $g$ and number of colours $N_c$ and factorized projectile
and target factors.

In Fig. \ref{fig4}  and all the following we show all partons with
solid lines. The upper and
bottom lines refer to the projectile and target quarks respectively.
Vertical lines refer to reggeons (just longitudinal gluons in our approximation).
The rest horizontal lines describe interaction kernels $V$ and $W$ and
correspond to real transverse gluons.
The three lines of  diagrams in Fig. \ref{fig4} correspond to the three
terms in Fig. \ref{fig2}. To the first and last lines one should add
similar diagrams which make the contribution symmetric in the two lower
legs.

To write down the corresponding analytic experssions we denote the
observed gluon transverse momentum and rapidity as $k_1$ and $y_1$ and
those of the softer gluon $k_2$ and $y_2$ with $y_2<<y_1$ and
$k_{2+}<<k_{1+}$.
The transverse momenta of the final reggeons  coupled to
the first target quark are denoted $q_1$ and $q_2$. For the inclusive
cross-section on a heavy nucleus the total momentum transferred to the
nucleus components is small. So we take $q_1+q_2=0$. The transverse
momenta of the second pair of final reggeons are $q_3$ and $q_4$ with
$q_3+q_4=0$. One is to perform integrations over  momenta of the
unobserved emitted gluon $k_2$ and final quark lines $q_1$ and $q_2$.
We omit the factors which arise from the final quark lines, so that
the final factor which appears in the inclusive cross-section is
\beq
2\alpha_sN_c\int\frac{dk_{2+}}{k_{2+}}\frac{d^2k_2}{4\pi^2}
\frac{d^2q_1}{(2\pi)}
\frac{d^2q_4}{(2\pi)}.
\label{ink2}
\eeq
This factor will also be suppressed
and assumed to be added to our expressions in the
end.
One has also to take into account that the pomeron wave function vanishes
when the two reggeons are located at the same point. In the momentum space
it means that integrated over its relative momentum the pomeron vanishes.
As a result all contributions which do not depend either on $q_1$ or
on $q_4$ vanish after integrations over $q_1$ and $q_4$. This allows to drop
diagrams like Fig \ref{fig4},3.

In terms of $V$, $W$ and $\omega$ the NLO contribution
to the KT cross-section is given by the sum
\beq
\frac{8\pi^3d\sigma^{KT}}{dyd^2k_1d^2b}\equiv
I^{KT}=\sum_{i=1}^3I^{KT}_i,
\label{incdef}
\eeq
where the three terms $I^{KT}_i$, $i=1,2,3$, correspond to contributions
from diagrams in Fig. \ref{fig2}, and are given by
(see Appendix for some details)
\[ I_1^{KT}=2V(k_2,-k_2|k_2+k_1,-k_2-k_1)W(q_1,-q_{14},q_4|k_2,-k_2)\]\beq
+2\omega(q_{14})\Big(V(q_1,-q_1|q_1+k_1,-q_1-k_1)+
V(-q_4,q_4|-q_4+k_1,q_4-k_1)\Big),
\label{ikt1}
\eeq
\[I_2^{KT}=-V(q_{14},-q_{14}|q_{14}+k_2,-q_{14}-k_2)
V(q_{14}+k_2,-q_{14}-k_2|q_{14}+k_2+k_1,-q_{14}-k_2-k_1)\]\beq
+2\omega(q_{14})V(q_{14},-q_{14}|q_{14}+k_1,-q_{14}-k_1),
\label{ikt2}
\eeq
\[I_3^{KT}=
V(q_{14}+k_2,-q_{14}-k_2|q_{14}+k_2+k_1,-q_{14}-k_2-k_1)
\Big(-V(q_1,-q_1|q_1+k_2,-q_1-k_2)\]\[
-V(q_4,-q_4|q_4+k_2,-q_4-k_2)\Big)\]\beq
+2V(q_{14},-q_{14}|q_{14}+k_1,-q_{14}-k_1)
\Big(\omega(q_1)+\omega(q_4)\Big),
\label{ikt3}
\eeq
where $q_{14}=q_1+q_4$.
As indicated these expression are to be taken with a common factor
(\ref{ink2}).

For future comparison with the BB picture it is convenient to reexpress
these contributions via function
\beq F(q,p)\equiv(q|p)=\frac{q^2}{p^2(q-p)^2}=(q|q-p)=(-q|-p).
\label{deff}
\eeq
Then one has
\beq
V(q_1,-q_1|q_1+k_2,-q_1-k_2)= -2(q_1,k_2),
\eeq
\beq
\omega(q_1)=-\frac{1}{2}(q_1|k_2)
\eeq
and the NLO contributions to the KT cross-section become
\beq
I_1^{KT}=-4(k_2|-k_1)(q_{14}|q_1-k_2)+2(q_{14}|k_2)
\Big((q_1|-k_1)+(q_4|k_1)\Big),
\label{ikt11}
\eeq
\beq
I_2^{KT}=-2(q_{14}|-k_2)\Big((2(q_{14}+k_2|-k_1)-(q_{14}|-k_1)\Big),
\label{ikt21}
\eeq
\beq
I_3^{KT}=-2\Big((q_1|-k_2)+(q_4|-k_2)\Big)\Big(2(q_{14}+k_2|-k_1)-
(q_{14}|-k_1)\Big).
\label{ikt31}
\eeq
It is understood that in these expressions $k_1$ is fixed, $k_2$ is integrated
over as indicated in (\ref{ink2}) and $q_1$ and $q_2$ are integrated
over with  the pomeron wave functions in the lowest approximation.
It is these expressions that we shall compare with the BB cross-section.

\section{BB cross-section, generalties}

In the BB approach the inclusive gluon production is derived from the
contribution to the total cross-section by fixing the momentum and
rapidity of the observed gluon in the intermediate state.
The total cross-section on two centers is given by a set of all
reggeon diagrams connecting the projectile with the two targets and
including interaction between reggeons either by the Lipatov
kernel $V$, conserving their number, or by the Bartels kernel $W$,
increasing their number.

Comparison with the KT approach requires to take identical weights in the
momentum integrations and also  appropriate normalizations
of impact factors for the projectile and the two targets in the
triple pomeron contribution. The
weight adopted in the BB formalism is $1/(8\pi^3)$ different from the
standard one $1/(4\pi^2)$ for the Fourier transform.
Taking into account that each momentum integration but one is accompanied
by factor $g^2$ we can pass to the standard weight $1/(4\pi^2)$ substituting
\beq
 g^2\to {\tilde g}^2=\frac{g^2}{2\pi}=2\alpha_s.
\label{gtilde}
 \eeq
(The lifted extra integration with weight $1/(8\pi^3)$ leads to the
inclusive cross-section normalized as in (\ref{incdef})).
In the following substitution (\ref{gtilde}) is assumed to be made.
To properly normalize the impact factors in the triple pomeron
contribution we take into account the relation
between the functions $\Phi$ and $\Psi$ which give the sum of fan diagrams
and also between the projectile
impact factors $\chi$ and $D$ in the KT and BB approaches correspondingly
\footnote{This relation was first established in ~\cite{BLV}, where
however the correct weight factor in the momentum integrations
was not taken into account
nor the fact that the triple pomeron vertex for the
amplitude was twice smaller and of the opposite sign.}
\beq
 \Phi=-\tilde {g}^2\Psi,\ \ \chi=-\frac{1}{\tilde {g}^2}D.
 \label{phipsi}
\eeq
Suppression of all impact factors in the KT approach then translates into
division of the contribution of BB triple pomeron diagrams by
$1/{\tilde g}^2$ (the minus sign goes due to the fact that the diagrams
themselves refer to the triple cut contribution and should be taken
with the opposite sign for the amplitude). As a result the final
expressions will be accompanied by the same factor (\ref{ink2}).

All the diagrams can be separated
by the initial number of reggeons to describe transitions from 2 to 4
reggeons, 3 to 4 reggeons and 4 to 4 reggeons. In the NLO
the diagrams can include either two interactions ($V$ or $W$)
or one interaction and an insertion of the gluon Regge trajectory $\omega$,
which transforms the gluon into reggeon. In the last case the resulting
contribution gives a part of terms which correspond to the BFKL evolution of
the final pomerons, the other part given by the BFKL interaction $V$ inside
the pair of legs. The sum of all such terms can be calculated in the
straightforward manner noting that in the LO the sum of all BB
diagrams gives exactly the KT cross-section, that is twice the contribution
from Fig. \ref{fig3}. In the NLO the BFKL evolution of the legs
will give twice the contribution $I_3^{KT}$.
\beq
I_1^{BB}=2I_3^{KT}.
\label{ibb1}
\eeq

The rest diagrams can in their turn be split into two sets depending on
how the 4 final gluon legs are coupled to the two target quarks.
The first possibility is that counting from the left the first pair of gluons
(momenta $q_1$ and $q_2$) are coupled to the first target quark and the
second pair (momenta $q_3$ and $q_4$) are coupled to the second target quark.
We call such configuration diffractive. An alternative configuration is when
the first and last gluons (momenta $q_3$ and $q_4$) are coupled to the second
target quark and the middle pair (momenta $q_1$ and $q_2$) are coupled to the
first quark. We call this configuration 'double cut'
for reasons to be explained presently.
The third possible configuration (not
neighbouring gluons coupled to each target quark) is damped in the heavy
nucleus.

The inclusive cross-section corresponds to fixing the
observed gluon in the intermediate state (that is in some interaction $V$
or $W$). The result is to be interpreted as a product of a certain production
amplitude by another production amplitude taken complex conjugate.
This corresponds to cutting the diagram across the real intermediate states.
In both configurations this cut may be done in two different ways.
In the diffractive configuration the cut either does not pass through the
target quarks at all, the diffractive (D) contribution,
which corresponds to the physical diffractive case,
or passes through one of the target quarks,
single diffractive (SD) contribution. One can show that the D and SD
contribution of the same diagram enter with the opposite sign.
In the double cut contribution the cut can pass either through both the
target quark (DC contribution, hence the name) or again through only one
of the target quarks (single double cut (SDC) contribution). The DC
contribution enters with factor 2 and the SDC contribution with factor $-1$.

The final factor for each contribution is determined taking into account
the sum over colours, which can be taken directly from the diagram
noting that each 3 gluon vertex contributes $f^{abc}$ with gluon colours
$a,b$ and $c$ taken anticlockwise.

\begin{figure}
\hspace*{3 cm}
\epsfig{file=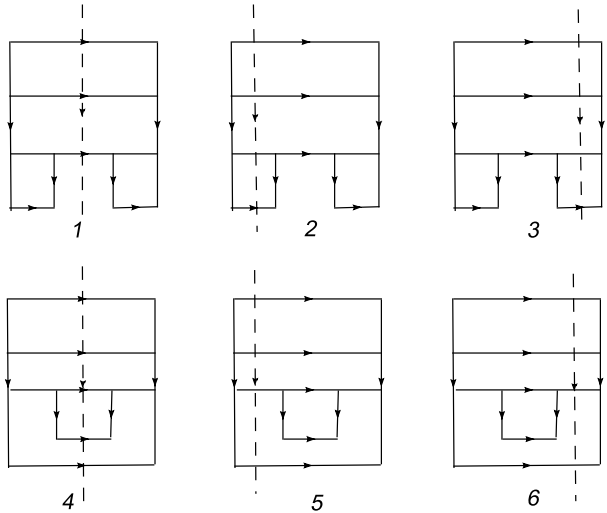,width=8 cm}
\caption{The diagrams in the BB approach which correspond to Fig
\ref{fig4},1 in the diffractive (1,2,3) and double cut (4,5,6)
configurations. (See notations in Fig. \ref{fig4}.)}
\label{fig5}
\end{figure}

To see how these rules work consider the diagrams of the same type as
Fig. \ref{fig4},1 for the KT amplitude in the preceding section. In the
BB framework one finds a set of diagrams shown in Fig. \ref{fig5}.

We recall that  all partons are shown by simple solid lines. The upper and
bottom lines refer to the projectile and target quarks respectively.
Vertcal lines refer to reggeons (just longitudinal gluons in our approximation).
The rest horizontal lines describe interaction kernels $V$ and $W$ and
correspond to real transverse gluons.
Colour summation for the  diagram gives factor $-N_c^5$, or just $-1$
suppressing the common $N_c^5$.

The DC contribution cancels with two SDC contributions.
The sign of the D contribution is reversed  by the two SD contributions.
 Writing the  $D$ contribution
in terms of $V$ and $W$ and reversing the sign we get  the term
corresponding to the
contribution of the diagram in Fig. \ref{fig4},1
for the KT cross-section but with a twice smaller coefficient
\beq
I_2^{BB}=-2(k_2|-k_1)(q_{14}|q_1-k_2).
\eeq

We are left with the sum  of all BB diagrams $I_R^{BB}$
excluding Fig. \ref{fig5} and all diagrams with
$\omega$ insertions and inserions of $V$ between the gluon legs connected
to a given target quark. They will be studied in the next two sections.

\section{Diffractive $4\to 4$ diagrams (case 1)}
All 16 diagrams cut in the middle are shown in Fig. \ref{fig5}.
\begin{figure}
\hspace*{1 cm}
\epsfig{file=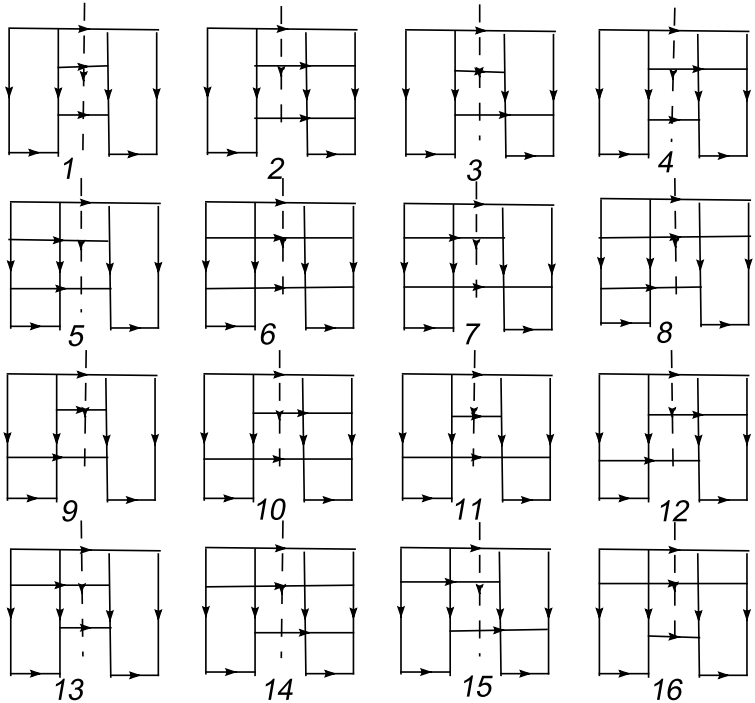,width=10 cm}
\caption{Feynman diagrams $4\to 4$ for the inclusive cross-section
in the diffractive configuration. (See notations in Fig. \ref{fig4}.)}
\label{fig6}
\end{figure}
One can split the whole set in three subsets depending on the position of
the upper  horizontal line corresponding to the observed gluon with the
larger rapidity.

The first set is formed by the diagrams (1),(3),(9)
and (11). Here the upper gluon connects the two inner reggeized gluons.
As a result there will be no contribution from the same diagrams with
one of the target quarks cut, the single diffractive (SD) contribution,
since such a cut does not contain the observed gluon.
Therefore contributions from this set remain as such.

The second set is formed by the diagrams (6),(8), (14) and (16).
Here the upper gluon connects the two outside reggeized gluons. So
it will be present in both SD cuts passing through the left and right target
quarks. Since each SD contribution is equal to the middle cut contribution
but has the opposite sign, in the sum  the contribution from this set
will change sign.

Finally we have the third set of diagrams  (2),(5), (4),(7),(10),
(12), (13), and (15) in which the observed gluon appears only in one
of the two SD contributions. So the contribution from this set cut in the
middle will be completely cancelled by the SD contribution.

Now we are going to demonstrate that the contributions from the
first and second set cancel in the sum. To do this we have to write
explicitly the integrand in $k_2$, $q_1$ and $q_4$ in each of the
relevant diagrams. Taking into account their  colour factors
we find the following expressions for the diagrams from the 1st set.
\[(1)=-\frac{1}{8}V(q_2,q_3|q_2+k_2,q_3-k_2)
V(q_2+k_2,q_3-k_2|q_2+k_2+k_1,q_3-k_2-k_1),\]
\[(3)=+\frac{1}{8}V(q_2,q_4|q_2+k_2,q_4-k_2)
V(q_2+k_2,q_3|q_2+k_2+k_1,q_3-k_1),\]
\[(9)=+\frac{1}{8}V(q_1,q_3|q_1+k_2,q_3-k_2)
V(q_2,q_3-k_2|q_2+k_1,q_3-k_2-k_1),\]
\[(11)=-\frac{1}{8}V(q_1,q_4|q_1+k_2,q_4-k_2)
V(q_2,q_3|q_2+k_1,q_3-k_1),\]
where in fact $q_2=-q_1$ and $q_3=-q_4$.

The diagrams in the second set have the integrands
\[(6)=-\frac{1}{8}V(q_1,q_4|q_1+k_2,q_4-k_2)
V(q_1+k_2,q_4-k_2|q_1+k_2+k_1,q_4-k_2-k_1),\]
\[(8)=+\frac{1}{8}V(q_1,q_3|q_1+k_2,q_3-k_2)
V(q_1+k_2,q_4|q_1+k_2+k_1,q_4-k_1),\]
\[(14)=+\frac{1}{8}V(q_2,q_4|q_2+k_2,q_4-k_2)
V(q_1,q_4-k_2|q_1+k_1,q_4-k_2-k_1),\]
\[(16)=-\frac{1}{8}V(q_2,q_3|q_2+k_2,q_3-k_2)V(q_1,q_4|q_1+k_1,q_4-k_1).\]
We observe that (6), (8), (14) and (16) are obtained from
(1), (3), (9) and (11) respectively by the substitutions
$q_1\leftrightarrow q_2$ and $q_3\leftrightarrow q_4$. However these
substitutions do not change the result after integration over $q_1$ and
$q_4$, so that the diagrams have in fact the same values. However, as
mentioned, the SD contribution in fact changes sign of the diagrams in the
second set. So in the sum they cancel with the first set.

\begin{figure}
\hspace*{2 cm}
\epsfig{file=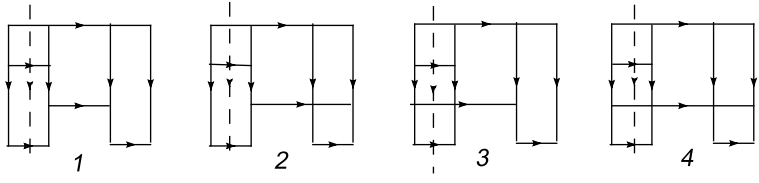,width=10 cm}
\caption{Single cut  diagrams in the diffractive configuration.
To the shown diagrams complex conjugate ones are to be added.
(See notations in Fig. \ref{fig4}.)}
\label{fig7}
\end{figure}

Thus the $4\to 4$
diagrams shown in Fig. \ref{fig6} cancel in the sum and give no
contribution to the inclusive cross-section. Additional  8 diagrams in the
SD configuration are shown in Fig. \ref{fig7} plus their complex conjugate.

The corresponding integrands are
\[ (1)=-\frac{1}{8}V(q_2,q_3|q_2+k_2,q_3-k_2)
V(q_1,q_2+k_2|q_1+k_1,q_2+k_2-k_1),\]
\[ (2)=+\frac{1}{8}V(q_2,q_4|q_2+k_2,q_4-k_2)
V(q_1,q_2+k_2|q_1+k_1,q_2+k_2-k_1),\]
\[ (3)=+\frac{1}{8}V(q_1,q_3|q_1+k_2,q_3-k_2)
V(q_1+k_2,q_2|q_1+k_2+k_1,q_2-k_1),\]
\[ (4)=-\frac{1}{8}V(q_1,q_4|q_1+k_2,q_4-k_2)
V(q_1+k_2,q_2|q_1+k_2+k_1,q_2-k_1).\]
The change $q_3\lra q_4$ shows that (1)+(2)=(3)+(4)$=0$.
So the conclusion is that all $4\to 4$ diffractive contribution cancels out.

To accommodate our notations to cover all cases we present our results in the
form
\[ V(q_1,q_2|q_1+k_2,q_4-k_2)\cdot X\]
for diffractive and  double cut contributions, to be taken with coefficient
$+1$ in
the total cross-section and
\[ V(q_1,q_2|q_1+k_2,q_4-k_2)\cdot Y\]
for all single cut contributions, to be taken with coefficient $-1$ in the
total cross-section.
We have found that in case 1
\beq
X_1=Y_1=0.
\eeq

\section{Diffractive $2\to 4$ diagrams (case 2)}
\begin{figure}
\hspace*{1 cm}
\epsfig{file=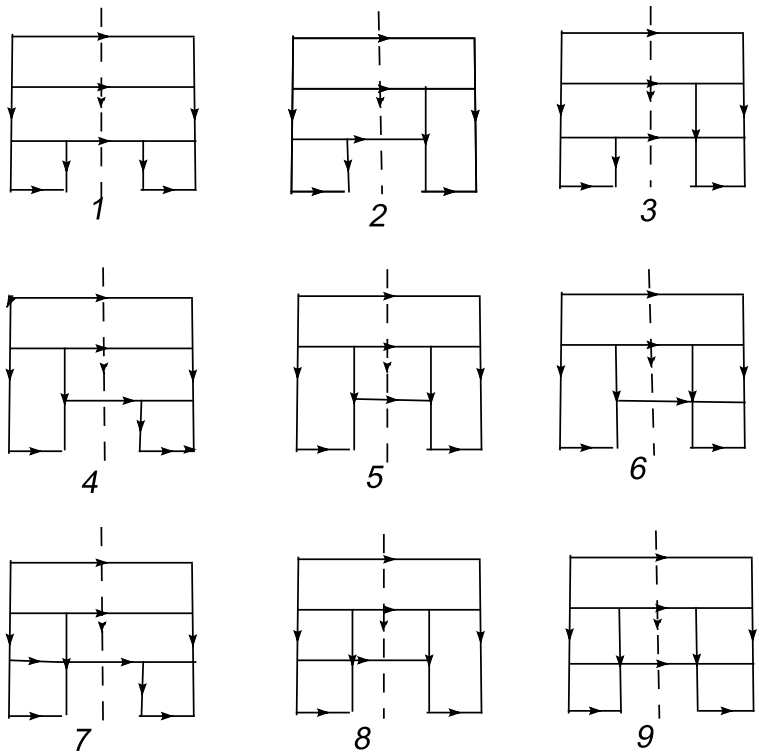,width=10 cm}
\caption{Feynman diagrams $2\to 4$ for the inclusive cross-section
in the diffractive configuration. (See notations in Fig. \ref{fig4}.)}
\label{fig8}
\end{figure}

Diagrams $2\to 4$ are
shown in Fig. \ref{fig8}, cut in the middle. One finds that diagrams
(2) and (3)
are cancelled with one of the SD contributions, that with the cut left target
quark. Diagrams (4) and (7) are cancelled with the SD contribution
corresponding to the cut right target quark. Diagram (1), which summed
with the two SD contributions changes its sign, has already been studied in
Sec. 3.  So we are left with diagrams (5), (6), (8), (9) and 12
SD diagrams.

The integrands for diagrams (5), (6), (8) and (9) are
\[ (5)=-\frac{1}{4}V(q_2,q_3|q_2+k_2,q_3-k_2)
W(q_1,q_2+q_3,q_4|q_1+q_2+k_1+k_2,q_4+q_3-k_1-k_2),\]
\[ (6)=+\frac{1}{4}V(q_2,q_4|q_2+k_2,q_4-k_2)
W(q_1,q_2+q_3+k_2,q_4-k_2|q_1+q_2+k_1+k_2,q_4+q_3-k_1-k_2),\]
\[ (8)=+\frac{1}{4}V(q_1,q_3|q_1+k_2,q_3-k_2)
W(q_1+k_2,q_2+q_3-k_2,q_4|q_1+q_2+k_1+k_2,q_4+q_3-k_1-k_2),\]
\[ (9)=-\frac{1}{4}V(q_1,q_4|q_1+k_2,q_4-k_2)
W(q_1+k_2,q_2+q_3,q_4-k_2|q_1+q_2+k_1+k_2,q_4+q_3-k_1-k_2).\]
Using  invariance under substitutions $q_1\leftrightarrow q_2$
and $q_3\leftrightarrow q_4$ we transform the Lipatov kernel in the first three term
to the common form $V(q_1,q_4|q_1+k_2,q_4-k_2)$.
In the sum of these diagrams we then get
\beq
D_{2\to 4}= (5)+(6)+(8)+(9)=V(q_1,q_4|q_1+k_2,q_4-k_2)\cdot X_2,
\eeq
where
\[
X_2=
-\frac{1}{4}W(q_2,q_1+q_4,q_3|q_1+q_2+k_1+k_2,q_3+q_4-k_1-k_2)\]\[
+\frac{1}{4}W(q_2,q_1+q_3+k_2,q_4-k_2|q_1+q_2+k_1+k_2,q_3+q_4-k_1-k_2)\]\[
+\frac{1}{4}W(q_1+k_2,q_2+q_4-k_2,q_3|q_1+q_2+k_1+k_2,q_3+q_4-k_1-k_2)\]\beq
-\frac{1}{4}W(q_1+k_2,q_2+q_3,q_4-k_2|q_1+q_2+k_1+k_2,q_3+q_4-k_1-k_2).
\eeq

\begin{figure}
\hspace*{1 cm}
\epsfig{file=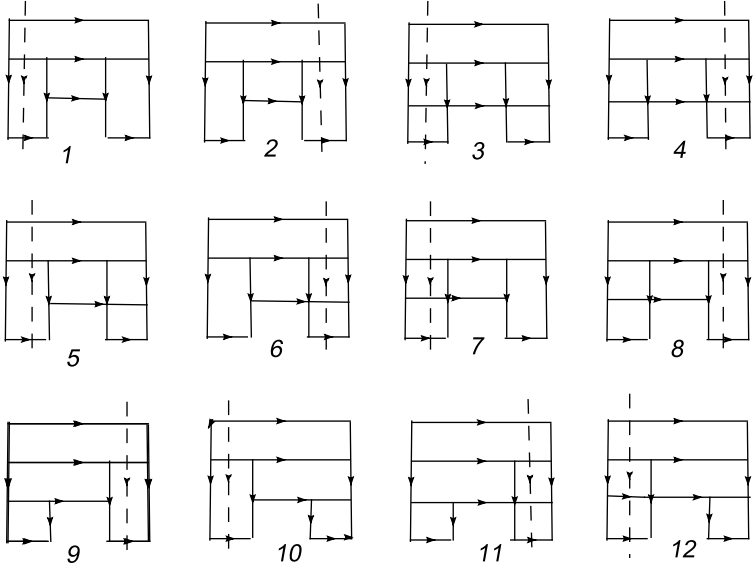,width=10 cm}
\caption{Feynman diagrams $2\to 4$ for the inclusive cross-section
in the diffractive configuration with one of the target quarks cut.
(See notations in Fig. \ref{fig4}.)}
\label{fig9}
\end{figure}

Now we consider the SD terms. The corresponding 12 diagrams are
presented in Fig.\ref{fig9}.
The integrands are
\[ (1)=(2)=-\frac{1}{4}V(q_2,q_3|q_2+k_2,q_3-k_2)
W(q_1,q_2+q_3,q_4|q_1+k_1,q_2+q_3+q_4-k_1),\]
\[ (5)= (8)=+\frac{1}{4}V(q_2,q_4|q_2+k_2,q_4-k_2)
W(q_1,q_2+q_3+k_2,q_4-k_2|q_1+k_1,q_2+q_3+q_4-k_1),\]
\[ (6)=(7)=+\frac{1}{4}V(q_2,q_4|q_2+k_2,q_4-k_2)
W(q_1,q_2+q_3+k_2,q_4-k_2|q_1+q_2+q_3+k_1+k_2,q_4-k_2-k_1),
\]
\[ (3)=(4)=-\frac{1}{4}V(q_1,q_4|q_1+k_2,q_4-k_2)
W(q_1+k_2,q_2+q_3,q_4-k_2|q_1+k_1+k_2,q_2+q_3+q_4-k_1-k_2),\]
\[ (9)=(10)=-\frac{1}{2}W(q_1,q_2,q_3|q_1+q_2+k_2,q_3-k_2)
W(q_1+q_2+k_2,q_3-k_2,q_4|q_1+q_2+q_3+k_1,q_4-k_1),\]
\[ (11)=(12)=+\frac{1}{2}W(q_1,q_2,q_4|q_1+q_2+k_2,q_4-k_2)\]\[
W(q_1+q_2+k_2,q_3,q_4-k_2|q_1+q_2+q_3+k_1+k_2,q_4-k_2-k_1).\]

In (1), (2), $(5)-(8)$ we again transform the $V$-factor to the
common form $V(q_1,q_4|q_1+k_2,q_4-k_2)$ using the
symmetry under  reflection $q_1\to -q_1$, $q_4\to -q_4$.
In (9)-(12) we use $q_1+q_2=0$ and the identity
\beq
W(q_1,-q_1,q_3|k_2,q_3-k_2)=-V(-q_1,q_3|-q_1+k_2,q_3-k_2).
\label{identity}
\eeq
Then we get
\[ (9)=(10)=
\frac{1}{2}V(q_1,q_4|q_1+k_2,q_4-k_2)
W(k_2,q_4-k_2,q_3|q_4+k_1,q_3-k_1)\]
and
\[ (11)=(12)=-
\frac{1}{2}V(q_1,q_4|q_1+k_2,q_4-k_2)
W(k_2,q_3,q_4-k_2|q_3+k_1+k_2,q_4-k_2-k_1).
\]

So we find
\beq
SD_{2\to 4}=\sum_{j=1}^{12} (j)=V(q_1,q_4|q_1+k_2,q_4-k_2)\cdot Y_2,
\eeq
where
\[
Y_2=
-\frac{1}{2}W(q_2,q_1+q_4,q_3|q_2+k_1,q_1+q_3+q_4-k_1)\]\[
-\frac{1}{2}W(q_1+k_2,q_2+q_3,q_4-k_2|q_1+k_1+k_2,q_2+q_3+q_4-k_1-k_2)\]\[
+\frac{1}{2}W(q_2,q_1+q_3+k_2,q_4-k_2|q_2+k_1,q_1+q_3+q_4-k_1)\]\[
+\frac{1}{2}W(q_2,q_1+q_3+k_2,q_4-k_2|q_1+q_2+q_3+k_1+k_2,q_4-k_1-k_2)\]\beq
+W(k_2,q_4-k_2,q_3|q_4+k_1,q_3-k_1)
-W(k_2,q_3,q_4-k_2|q_3+k_1+k_2,q_4-k_1-k_2).
\eeq

\section{Diffractive $3\to 4$ diagrams (case 3)}

\begin{figure}
\hspace*{1 cm}
\epsfig{file=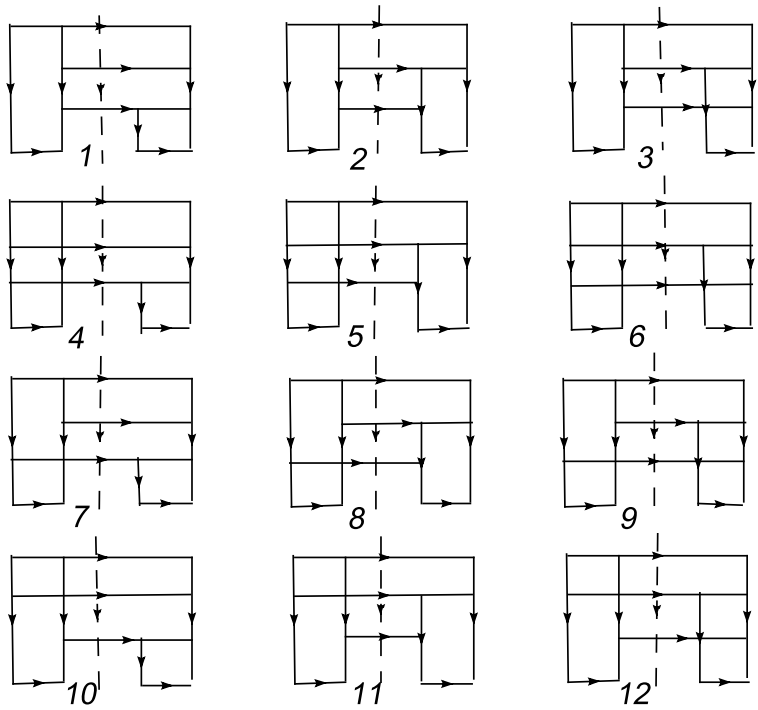,width=10 cm}
\caption{Feynman diagrams $3\to 4$ for the inclusive cross-section
in the diffractive configuration. To the shown diagrams one should add
their complex conjugate ones. (See notations in Fig. \ref{fig4}.)}
\label{fig10}
\end{figure}

The 24 $3\to 4$ diagrams are shown in Fig. \ref{fig10}
plus their complex conjugate.

One observes that the contributions from  diagrams (1), and (7)
are cancelled by the SD contributions coming from the cut right target quark
and from  diagrams (5), (6), (11) and (12) by the SD contribution
from the cut left target quark. Diagrams (4) and (10) change sign by the
two SD contributions from the cut left and right target quarks. Diagrams
(2), (3), (8) and (9) remain as such.

Integrands for the remaining  diagrams
are
\[ (2)=-\frac{1}{8}V(q_2,q_3|q_2+k_2,q_3-k_2)
W(q_2+k_2,q_3-k_2,q_4|q_2+k_1+k_2,q_3+q_4-k_1-k_2),\]
\[ (3)=\frac{1}{8}V(q_2,q_4|q_2+k_2,q_4-k_2)
W(q_2+k_2,q_3,q_4-k_2|q_2+k_1+k_2,q_3+q_4-k_1-k_2),\]
\[ (8)=\frac{1}{8}V(q_1,q_3|q_1+k_2,q_3-k_2)
W(q_2,q_3-k_2,q_4|q_2+k_1,q_3+q_4-k_1-k_2),\]
\[ (9)=-\frac{1}{8}V(q_1,q_4|q_1+k_2,q_4-k_2)
W(q_2,q_3,q_4-k_2|q_2+k_1,q_3+q_4-k_1-k_2),
\]
\[ (4)=-\frac{1}{4}W(q_1,q_3,q_4|q_1+k_2,q_3+q_4-k_2)
V(q_1+k_2,q_3+q_4-k_2|q_1+k_1+k_2,q_3+q_4-k_1-k_2),\]
\[ (10)=\frac{1}{4}W(q_2,q_3,q_4|q_2+k_2,q_3+q_4-k_2)
V(q_1,q_3+q_4-k_2|q_1+k_1,q_3+q_4-k_1-k_2).\]

For (4) and (10) we again use Eq.(\ref{identity})
and changing $q_4\to -q_4$  find
\[(4)=\frac{1}{4}V(q_1,q_4|q_1+k_2,q_4-k_2)
V(q_1+k_2,-k_2|q_1+k_1+k_2,-k_1-k_2),\]
\[(10)=-\frac{1}{4}V(q_1,q_4|q_1+k_2,q_4-k_2)
V(q_2,-k_2|q_2+k_1,-k_1-k_2).\]

Separating factor $V(q_1,q_4|q_1+k_2,q_4-k_2)$ in the rest of diagrams
and doubling  we obtain the total diffractive contribution from $3\to 4$
transitions
\beq
D_{3\to 4}=V(q_1,q_4|q_1+k_2,q_4-k_2)\cdot X_3,
\eeq
where
\[
X_3=-\frac{1}{4}W(q_1+k_1,q_4-k_2,q_3|q_1+k_1+k_2,-k_1-k_2)\]\[
+\frac{1}{4}W(q_1+k_1,q_3,q_4-k_2|q_1+k_1+k_2,-k_1-k_2)\]\[
+\frac{1}{4}W(q_2,q_4-k_2,q_3|q_2+k_1,-k_1-k_2)
-\frac{1}{4}W(q_2,q_3,q_4-k_2|q_1+k_1,-k_1-k_2)\]\beq
+\frac{1}{2}V(q_1+k_2,-k_2|q_1+k_1+k_2,-k_1-k_2)
-\frac{1}{2}V(q_2,-k_2|q_2+k_1,-k_1-k_2).
\eeq

\begin{figure}
\hspace*{1 cm}
\epsfig{file=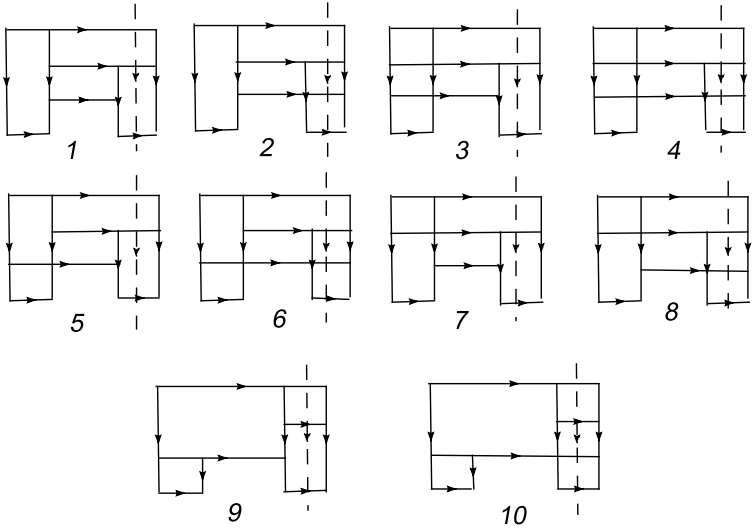,width=10 cm}
\caption{Feynman diagrams $3\to 4$ for the inclusive cross-section
in the diffractive configuration with one of the target quarks cut.
(See notations in Fig. \ref{fig4}.)}
\label{fig11}
\end{figure}

Now we pass to SD diagrams which remain after cancellation with the
diffractive diagrams. They are 10 and are shown in Fig. \ref{fig11}.
The integrands are
\[ (1)=-\frac{1}{8}V(q_2,q_3|q_2+k_2,q_3-k_2)
W(q_2+k_2,q_3-k_2,q_4|q_2+q_3+k_1,q_4-k_1),\]
\[ (2)=\frac{1}{8}V(q_2,q_4|q_2+k_2,q_4-k_2)
W(q_2+k_2,q_3,q_4-k_2|q_2+q_3+k_1+k_2,q_4-k_1-k_2),
\]
\[ (3)=\frac{1}{8}V(q_1,q_3|q_1+k_2,q_3-k_2)
W(q_1+k_2,q_3-k_2,q_4|q_1+q_3+k_1,q_4-k_1),
\]
\[ (4)=-\frac{1}{8}V(q_1,q_4|q_1+k_2,q_4-k_2)
W(q_1+k_2,q_3,q_4-k_2|q_1+q_3+k_1+k_2,q_4-k_1-k_2),
\]
\[ (5)=\frac{1}{8}V(q_1,q_3|q_1+k_2,q_3-k_2)
W(q_2,q_3-k_2,q_4|q_2+q_3+k_1-k_2,q_4-k_1),
\]
\[ (6)=-\frac{1}{8}V(q_1,q_4|q_1+k_2,q_4-k_2)
W(q_2,q_3,q_4-k_2|q_2+q_3+k_1,q_4-k_1-k_2),
\]
\[ (7)=-\frac{1}{8}V(q_2,q_3|q_2+k_2,q_3-k_2)
W(q_1,q_3-k_2,q_4|q_1+q_3+k_1-k_2,q_4-k_1),
\]
\[ (8)=\frac{1}{8}V(q_2,q_4|q_2+k_2,q_4-k_2)
W(q_1,q_3,q_4-k_2|q_1+q_3+k_1,q_4-k_1-k_2),
\]
\[ (9)=-\frac{1}{4}W(q_1,q_2,q_3|q_1+q_2+k_2,q_3-k_2)
V(q_3-k_2,q_4|q_3+k_1-k_2,q_4-k_1),
\]
\[ (10)=\frac{1}{4}W(q_1,q_2,q_4|q_1+q_2+k_2,q_4-k_2)
V(q_3,q_4-k_2|q_3+k_1,q_4-k_1-k_2).
\]

We observe that
\[(3)=-(1)\ \ {\rm with}\ \ q_1\lra q_2,\ \
(4)=-(2)\ \ {\rm with}\ \ q_1\lra q_2,\]
\[(7)=-(5)\ \ {\rm with}\ \ q_1\lra q_2,\ \
(8)=-(6)\ \ {\rm with}\ \ q_1\lra q_2.\]
Changing $q_3\lra q_4$ and $k_1\to -k_1$  in (10) we
also find $(9)=-(10)$.
So the sum of all 8 diagrams is zero and
the SD$_{3\to4}$ diagrams give no contribution : $Y_3=0$

\section{Double cut (DC) $4\to 4$ diagrams (case 4)}
\begin{figure}
\hspace*{1 cm}
\epsfig{file=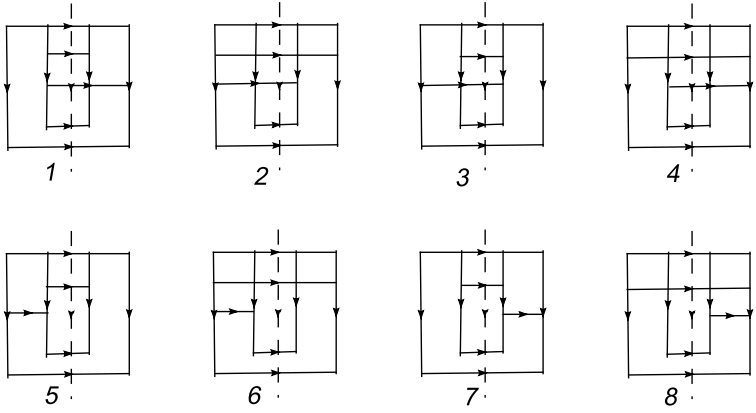,width=10 cm}
\caption{Feynman diagrams $4\to 4$ for the inclusive cross-section
in the double cut configuration. (See notations in Fig. \ref{fig4}.)}
\label{fig12}
\end{figure}

The eight $4\to 4$ middle cut DC diagrams with non-zero colour factors are shown
in Fig. \ref{fig12}.  Of them 4 diagrams (2), (4), (6) and (8) are
completely cancelled by the corresponding single cut contributions.
We are left with 4 diagrams (1), (3), (5) and (7).
Their direct calculations give the following
integrands for the integrals over $k_2$, $q_1$ and $q_4$
\[ (1)=(3)=2\times\Big(\frac{1}{8}\Big)V(q_1,q_4|q_1+k_2,q_4-k_2)
V(q_1+k_2,q_2|q_1+k_1+k_2,q_2-k_1),\]
\[ (5)=(7)=-2\times\Big(\frac{1}{8}\Big)V(q_2,q_4|q_2+k_2,q_4-k_2)
V(q_1,q_2+k_2|q_1+k_1,q_2+k_2-k_1),\]
where the separated factor two  comes from taking the double cut
uncompensated by single cut contributions. A change $q_1\lra q_2$ in the second
expression converts it into the first with the opposite sign.
So $(1)+(5)=0$ and the contribution from terms $4\to 4$
shown in Fig. \ref{fig12}
is completely cancelled as in the diffractive configuration.

\begin{figure}
\hspace*{4 cm}
\epsfig{file=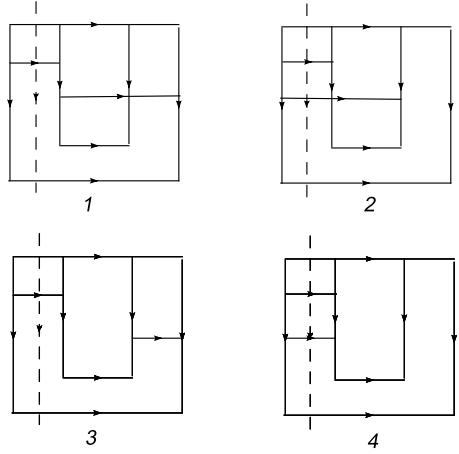,width=6 cm}
\caption{Single cut  diagrams in the double cut configuration.
To the shown diagrams their complex conjugate ones are to be added.
(See notations in Fig. \ref{fig4}.)}
\label{fig13}
\end{figure}

Additional  8 diagrams in the
SD configuration are  shown in Fig. \ref{fig13} plus their complex conjugate.

The corresponding integrands are
\[ (1)=\frac{1}{8}V(q_1,q_4|q_1+k_2,q_4-k_2)
V(q_3,q_1+k_2|q_3+k_1,q_1+k_2-k_1),\]
\[ (2)=\frac{1}{8}V(q_3,q_2|q_3+k_2,q_2-k_2)
V(q_3+k_2,q_1|q_3+k_1+k_2,q_1-k_1),
\]
\[(3)=-\frac{1}{8}V(q_2,q_4|q_2+k_2,q_4-k_2)
V(q_3,q_1|q_3+k_1,q_1-k_1),
\]
\[ (4)=-\frac{1}{8}V(q_3,q_1|q_3+k_2,q_1-k_2)
V(q_3+k_2,q_1-k_2|q_3+k_1+k_2,q_1-k_1-k_2).\]

To separate the common factor $V(q_1,q_4|q_1+k_2,q_4-k_2)$ we change
$q_1\lra q_2$, $q_3\lra q_4$ and $k_2\to -k_2$ in (2),
$q_2\lra q_1$ in (3) and $q_3\lra q_4$ and $k_2\to -k_2$ in (4).
Then we get
\beq
SDC_{4\to 4}=2\Big((1)+(2)+(3)+(4)\Big)=V(q_1,q_4|q_1+k_2,q_4-k_2)\cdot Y_4,
\eeq
where
\[Y_4=
\frac{1}{4}V(q_3,q_1+k_2|q_3+k_1,q_1+k_2-k_1)
+\frac{1}{4}V(q_4-k_2,q_2|q_4+k_1-k_2,q_2-k_1)\]\beq
-\frac{1}{4}V(q_3,q_2|q_3+k_1,q_2-k_1)
-\frac{1}{4}V(q_4-k_2,q_1+k_2|q_4+k_1-k_2,q_1-k_1+k_2).
\eeq

\section{DC $2\to 4$ diagrams (case 5)}
\begin{figure}
\hspace*{1 cm}
\epsfig{file=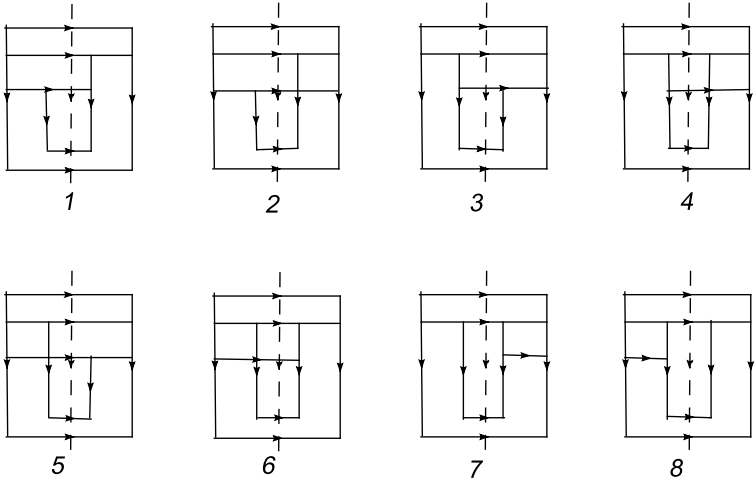,width=10 cm}
\caption{Feynman diagrams $2\to 4$ for the inclusive cross-section
in the double cut configuration. (See notations in Fig. \ref{fig4}.)}
\label{fig14}
\end{figure}

8 relevant diagrams are shown in Fig. \ref{fig14}.
Diagrams (1), (2), (3) and (5) have one corresponding
single cut contribution, so that their initial coefficient 2 is reduced to
unity. Diagrams (4), (6), (7) and (8) remain with the coefficient 2.
The integrands obtained from the straightforward calculations are
\[ (1)=(3)=-\frac{1}{2}W(q_3,q_1,q_2|q_3+k_2,q_1+q_2-k_2)\]\[
W(q_3+k_2,q_1+q_2-k_2,q_4|q_3+k_1+k_2,q_1+q_2+q_4-k_1-k_2),
\]
\[ (2)=(5)=\frac{1}{4}W(q_3,q_1,q_4|q_1+q_3+k_2,q_4-k_2)\]\[
W(q_1+q_3+k_2,q_2,q_4-k_2|q_1+q_3+k_1+k_2,q_2+q_4-k_1-k_2),
\]
\[ (4)=(6)=2\times\Big(\frac{1}{4}\Big)V(q_1,q_4|q_1+k_2,q_4-k_2)\]\[
W(q_3,q_1+q_2+k_2,q_4-k_2|q_1+q_3+k_1+k_2,q_2+q_4-k_1-k_2),
\]
\[ (7)=(8)=-2\times\Big(\frac{1}{4}\Big)V(q_2,q_4|q_2+k_2,q_4-k_2)\]\[
W(q_3,q_1+q_2+k_2,q_4-k_2|q_1+q_3+k_1,q_2+q_4-k_1).
\]

The first and last expressions can be transformed to our standard form with
factor $V(q_1,q_4,q_+k_2,q_4-k_2)$ using Eq.(\ref{identity}) and the change
$q_3\lra q_4$ and $k_2\to -k_2$
\[ (1)=(3)=\frac{1}{2}V(q_1,q_4|q_1+k_2,q_4-k_2)
W(q_4-k_2,k_2,q_3|q_4+k_1-k_2,q_3-k_1+k_2).\]
In the expression for (7) it is sufficient to change $q_1\lra q_2$.
Contribution (2)=(5) cannot be transformed to the standard form with
factor $V(q_1,q_4|q_1+k_2,q_4-k_2)$ separated. However we shall see that this
contribution is cancelled with a similar one coming from the single cut diagrams.

As a result the  DC contribution from $2\to 4$ transitions is
\beq
DC_{2\to 4}=(1)+(3)+(4)+(6)+(7)+(8)=V(q_1,q_4|q_1+k_2,q_4-k_2)\cdot X_5,
\eeq
where
\[X_5=W(q_4-k_2,k_2,q_3|q_4+k_1-k_2,q_3-k_1+k_2)\]
\[+W(q_3,q_1+q_2+k_2,q_4-k_2|q_1+q_3+k_1+k_2,q_2+q_4-k_1-k_2)\]\beq
-W(q_3,q_1+q_2+k_2,q_4-k_2|q_2+q_3+k_1,q_1+q_4-k_1).
\eeq

\begin{figure}
\hspace*{1 cm}
\epsfig{file=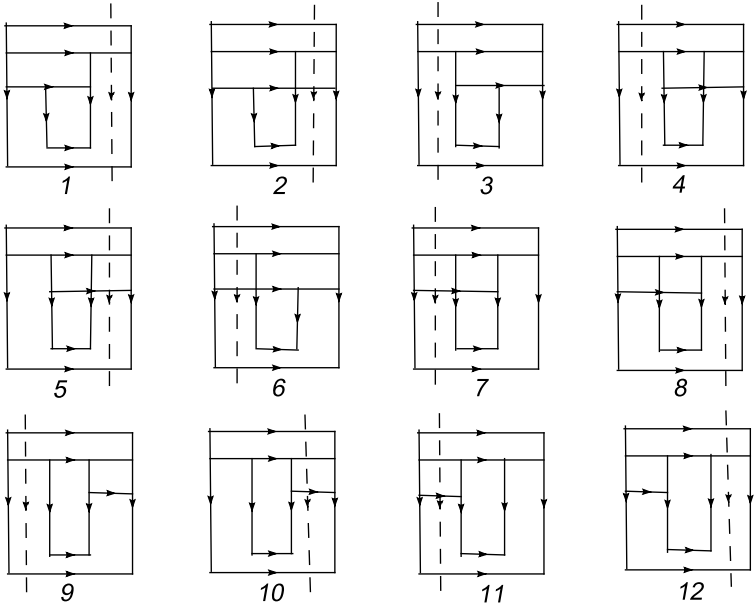,width=10 cm}
\caption{Feynman diagrams $2\to 4$ for the inclusive cross-section
with one of the target quarks cut associated with the double cut
configuration. (See notations in Fig. \ref{fig4}.)}
\label{fig15}
\end{figure}

Now we pass to single cut diagrams associated with the DC $2\to 4$ diagrams
(SDC $2\to 4$ diagrams). The 12 SDC $2\to 4$ diagrams are shown
in Fig. \ref{fig15}. They are pairwise equal, so we have to study 6
different contributions.
Calculations give the following integrands
\[(1)=(3)=-\frac{1}{2}W(q_3,q_1,q_2|q_3+k_2,q_1+q_2-k_2)
W(q_3+k_2,q_1+q_2-k_2,q_4|q_1+q_2+q_3+k_1,q_4-k_1),\]
\[(2)= (6)=\frac{1}{4}W(q_3,q_1,q_4|q_1+q_3+k_2,q_4-k_2)\]\[
W(q_1+q_3+k_2,q_2,q_4-k_2|q_1+q_2+q_3+k_1+k_2,q_4-k_1-k_2),\]
\[(4)= (8)=\frac{1}{4}V(q_1,q_4|q_1+k_2,q_4-k_2)
W(q_3,q_1+q_2+k_2,q_4-k_2|q_3+k_1,q_1+q_2+q_4-k_1),\]
\[(5)= (7)=\frac{1}{4}V(q_1,q_4|q_1+k_2,q_4-k_2)
W(q_3,q_1+q_2+k_2,q_4-k_2|q_1+q_2+q_3+k_1+k_2,q_4-k_1-k_2),\]
\[(9)= (12)=-\frac{1}{4}V(q_2,q_4|q_2+k_2,q_4-k_2)
W(q_3,q_1+q_2+k_2,q_4-k_2|q_3+k_1,q_1+q_2+q_4-k_1),
\]
\[(10)= (11)=-\frac{1}{4}V(q_2,q_4|q_2+k_2,q_4-k_2)\]\[
W(q_3,q_1+q_2+k_2,q_4-k_2|q_1+q_2+q_3+k_1+k_2,q_4-k_1-k_2).
\]

Changing $q_1\lra q_2$ we find that
$(4)+(9)=(5)+(10)=0$. So the total contribution reduces to
$2(1)+2(2)$. Contribution (1) can be transformed to the standard form.
Using (\ref{identity}) and changing $q_3\lra q_4$ and $k_2\to -k_2$ we find
\[(1)=(3)=\frac{1}{2}V(q_1,q_4|q_1+k_2,q_4-k_2)
W(q_4-k_2,k_2,q_3|q_4+k_1,q_3-k_1).\]

Now  we demonstrate that contributions from (2)+(6)
cancel similar contributions (2)+(5) from the DC$_{2\to 4}$ diagrams.
We make a change of the integration momentum $k_2\to -k_2-q_1$ in SDC(2)
to get
\[SDC(2)=-W(q_3,q_2,q_4|q_3-k_2,q_1+q_4+k_2)\]\[
W(q_3-k_2,q_2,q_1+q_4+k_2|q_2+q_3+k_1-k_2,q_1+q_4-k_1+k_2).\]
Change $q_3\lra q_4$ and $k_1\to -k_1$ transforms SDC(2) into DC(2).
Since they enter the inclusive cross-section with opposite signs,
their contributions cancel.

As a result the only contribution left from the single cut diagrams $2\to 4$ comes
from the sum (1)+(3)
\beq
SDC_{2\to 4}=(1)+(3)=V(q_1,q_4|q_1+k_2,q_4-k_2)\cdot Y_5,
\eeq
where
\beq
Y_5=W(q_4-k_2,k_2,q_3|q_4+k_1,q_3-k_1).
\eeq

\section{DC $3\to 4$ diagrams (case 6)}
\begin{figure}
\hspace*{1 cm}
\epsfig{file=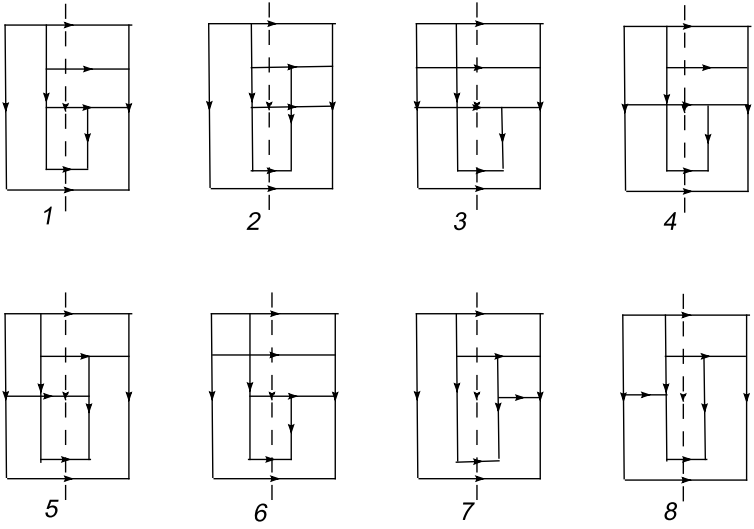,width=10 cm}
\caption{Feynman diagrams $3\to 4$ for the inclusive cross-section
in the double cut configuration. Complex conjugate diagrams are to be added.
(See notations in Fig. \ref{fig4}.)}
\label{fig16}
\end{figure}

The interference DC diagrams $3\to 4$ come in pairs. Half of them with
non-zero colour coefficients are shown in Fig. \ref{fig16} (8 diagrams).
Of them diagrams (3) and (6) are completely cancelled by the
associated single cut contributions. Diagrams (1) and (4) become half-wise
 cancelled
by the single cut contribution and its coefficient 2 is reduced to unity.
The rest 4 diagrams (2), (5), (7) and (8) remain with coefficient 2.
Direct calculation gives the following integrands
\[ (1)=-\frac{1}{4}W(q_1,q_2,q_4|q_1+q_2+k_2,q_4-k_2)
V(q_1+q_2+k_2,q_4-k_2|q_1+q_2+k_1+k_2,q_4-k_1-k_2),\]
\[ (2)=2\times\Big(\frac{1}{8}\Big)V(q_1,q_4|q_1+k_2,q_4-k_2)
W(q_1+k_2,q_2,q_4-k_2|q_1+k_1+k_2,q_2+q_4-k_1-k_2),\]
\[ (4)=\frac{1}{8}W(q_3,q_2,q_4|q_3+k_2,q_2+q_4-k_2)
V(q_1,q_2+q_4-k_2|q_1+k_1,q_2+q_4-k_1-k_2),\]
\[ (5)=2\times\Big(\frac{1}{8}\Big)V(q_3,q_2|q_3+k_2,q_2-k_2)
W(q_1,q_2-k_2,q_4|q_1+k_1,q_2+q_4-k_1-k_2),\]
\[ (7)=2\times\Big(-\frac{1}{8}\Big)V(q_2,q_4|q_2+k_2,q_4-k_2)
W(q_1,q_2+k_2,q_4-k_2|q_1+k_1,q_2+q_4-k_1),\]
\[ (8)=2\times\Big(-\frac{1}{8}\Big)V(q_3,q_1|q_3+k_2,q_1-k_2)
W(q_1-k_2,q_2,q_4|q_1+k_1-k_2,q_2+q_4-k_1).\]

As before we transform all expressions except
(4) to the standard form.
In (1) we use Eq.(\ref{identity}) to find
\[(1)=\frac{1}{4}V(q_1,q_4|q_1+k_2,q_4-k_2)
V(k_2,q_4-k_2|k_1+k_2,q_4-k_1-k_2).\]
In (5) we change $q_1\lra q_2$, $q_3\lra q_4$ and $k_2\to -k_2$.
In (7) we only change $q_1\lra q_2$. In (8) we change
$q_3\lra q_4$ and $k_2\to -k_2$.

Contribution from diagram (4) will be cancelled with a similar one
from the single cut diagram (6) (see below).

Combining all terms and doubling we get the total DC contribution
from $3\to 4$ diagrams
\beq
 DC_{3\to 4}=2\Big((1)+(2)+(5)+(7)+(8)\Big)=
V(q_1,q_4,q_1+k_2,q_4-k_2)\cdot X_6,
\eeq
where
\[X_6=
\frac{1}{2}V(k_2,q_4-k_2|k_1+k_2,q_4-k_1-k_2)\]
\[+\frac{1}{2}W(q_1+k_2,q_2,q_4-k_2|q_1+k_1+k_2,q_2+q_4-k_1-k_2)\]
\[+\frac{1}{2}W(q_2,q_1+k_2,q_3|q_2+k_1,q_1+q_3-k_1+k_2)\]\beq
-\frac{1}{2}W(q_2,q_1+k_2,q_4-k_2|q_2+k_1,q_1+q_4-k_1)
-\frac{1}{2}W(q_1+k_2,q_2,q_3|q_1+k_1+k_2,q_2+q_3-k_1).
\eeq

\begin{figure}
\hspace*{3 cm}
\epsfig{file=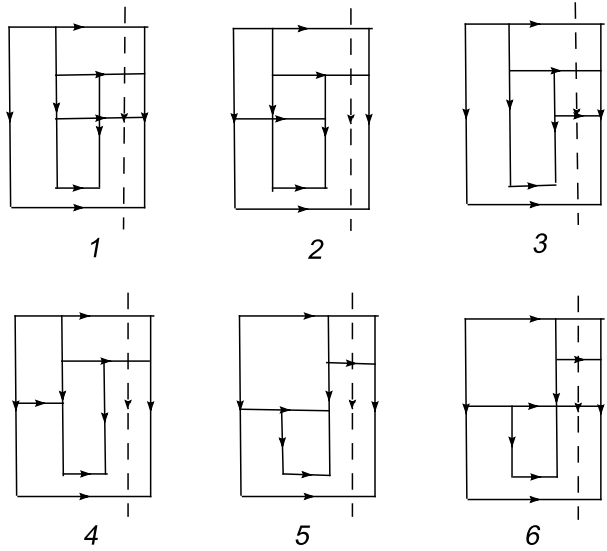,width=8 cm}
\caption{Feynman diagrams $3\to 4$ for the inclusive cross-section
with one of the target quarks cut associated with the double cut
configuration. Complex conjugate diagrams are to be added.
(See notations in Fig. \ref{fig4}.)}
\label{fig17}
\end{figure}

We now pass to the associated single cut diagrams (SDC for $3\to 4$).
Half of them with non-zero colour factors are shown in Fig. \ref{fig17}.
The integrands are
\[ (1)=\frac{1}{8}V(q_1,q_4|q_1+k_2,q_4-k_2)
W(q_1+k_2,q_2,q_4-k_2|q_1+q_2+k_1+k_2,q_4-k_1-k_2),\]
\[ (2)=\frac{1}{8}V(q_3,q_2|q_3+k_2,q_2-k_2)
W(q_1,q_2-k_2,q_4|q_1+q_2+k_1-k_2,q_4-k_1),\]
\[ (3)=-\frac{1}{8}V(q_2,q_4|q_2+k_2,q_4-k_2)
W(q_1,q_2+k_2,q_4-k_2|q_1+q_2+k_1+k_2,q_4-k_1-k_2),
\]
\[ (4)=-\frac{1}{8}V(q_3,q_1|q_3+k_2,q_1-k_2)
W(q_1-k_2,q_2,q_4|q_1+q_2+k_1-k_2,q_4-k_1),
\]
\[ (5^*)=-\frac{1}{4}W(q_1,q_2,q_4|q_1+q_2+k_2,q_4-k_2)
V(q_3,q_1+q_2+k_2|q_3+k_1,q_1+q_2+k_2-k_1),\]
\[ (6^*)=\frac{1}{8}W(q_3,q_2,q_4|q_3+k_2,q_2+q_4-k_2)
V(q_3+k_2,q_1|q_3+k_1+k_2,q_1-k_1).\]

As we shall see below (6) is cancelled with a similar contribution from DC.
The rest of contributions are easily transformed to the standard form.
In (5) we use (\ref{identity}) and change  $q_1\to -q_1$ to give
\[ (5)=\frac{1}{4}V(q_1,q_4|q_1+k_2,q_4-k_2)
V(q_3,k_2|q_3+k_1,k_2-k_1).\]
Next we change in (2) $q_1\lra q_2$, $q_3\lra q_4$ and $k_2\to -k_2$,
in (3) $q_1\lra q_2$ and in (4) $q_3\lra q_4$ and $k_2\to -k_2$.
We obtain
\beq
SDC_{3\to 4}=2\sum_{j=1}^5 (j)=V(q_1,q_4|q_1+k_2,q_4-k_2)\cdot Y_6,
\eeq
where
\[ Y_6=
 \frac{1}{4}W(q_1+k_2,q_2,q_4-k_2|q_1+q_2+k_1+k_2,q_4-k_1-k_2)\]
\[+\frac{1}{4}W(q_2,q_1+k_2,q_3|k_1+k_2,q_4-k_1)
-\frac{1}{4}W(q_2,q_1+k_2,q_4-k_2|k_1+k_2,q_4-k_1-k_2)\]\beq
-\frac{1}{4}W(q_1+k_2,q_2,q_3|k_1+k_2,q_3-k_1)
+\frac{1}{2}V(q_3,k_2|q_3+k_1,k_2-k_1).
\eeq

Finally we check that the contribution from diagram (6) cancels the
similar one DC(4) from
DC$_{3\to 4}$.
Transformation of the integration momentum $k_2\to q_2+q_4-q_3-k_2$ transforms
SDC(6) into DC(4). Since they enter the cross-section with opposite signs, they
cancel in the sum.

\section{Final expression for the BB inclusive cross-section}
It is convenient to express the found contributions $X_i$ and $Y_i$
in terms of function $F(q|p)\equiv(q|p)$ defined by (\ref{deff}).
Originally the sum of $X_i$ and $Y_i$ contains 114 functions $F$.
However many terms cancel within each $X_i$ and $Y_i$. After these
cancellations we find non-zero $X_i$ and $Y_i$ in the form

\[ X_2=
-\frac{1}{4}(q_{14}|q_1+k_1+k_2)
+\frac{1}{4}(q_1-q_4+k_2|q_1+k_1+k_2)\]
\beq
+\frac{1}{4}(-q_1+q_4-k_2|k_1-q_1)
-\frac{1}{4}+(-q_{14}|k_1-q_1).
\eeq

\[X_3=
\frac{1}{4}(q_{14}|q_1+k_1+k_2)
-\frac{1}{4}(q_1-q_4+k_2|q_1+k_1+k_2)\]
\[-\frac{1}{4}(-q_1+q_4-k_2|-q_1+k_1)
+\frac{1}{4}(-q_{14}|-q_1+k_1)\]
\[+\frac{1}{2}\Big((q_1|q_1+k_1+k_2)-(q_1+k_2|q_1+k_1+k_2)\Big)\]
\beq
-\frac{1}{2}\Big((-q_1-k_2|-q_1+k_1)-(-q_1|-q_1+k_1)\Big).
\eeq

\[Y_2=
-\frac{1}{2}\Big(-(q_4|-q_1+k_1)-(q_1|q_1-k_1)+(q_{14}|k_1)\]
\[-(-q_4+k_2|q_1+k_1+k_2)-(-q_1-k_2|-q_1-k_1-k_2)+
(-q_{14}|k_1)\]
\[+(-q_4+k_2|-q_1+k_1)+(q_1|q_1-k_1)-
(q_1-q_4+k_2|k_1)\]
\[+(-q_4+k_2|-q_4+k_1+k_2)+(q_1|q_4-k_1-k_2)-(q_1-q_4+k_2|-k_1)\Big)\]
\[-(q_4|q_4+k_1)-(-k_2|-q_4-k_1)+(q_4-k_2|-k_1)\]
\beq
+(k_2-q_4|-q_4+k_1+k_2)+(-k_2|q_4-k_1-k_2)-(-q_4|-k_1).
\eeq

\[X_5=
-(q_4|q_4+k_1-k_2)-(-q_4+k_2|-q_4-k_1+k_2)+(k_2|k_1)\]
\[-(-q_4+k_2|q_1-q_4+k_1+k_2)-(q_4|-q_1+q_4-k_1-k_2)\]
\beq
+(-q_4+k_2|-q_1-q_4+k_1)+(q_4|q_{14}-k_1).
\eeq

\[ X_6=\frac{1}{2}\Big[
(q_4|k_1+k_2)-(k_2|k_1+k_2)-(q_4-k_2|q_4-k_1-k_2)\]
\[+(q_4|q_1+k_1+k_2)-(-q_1+q_4-k_2|-q_1+q_4-k_1-k_2)\]
\[+(-q_4+k_2|-q_1+k_1)-(q_1-q_4+k_2|q_1-q_4-k_1+k_2)\]
\[-(q_4|-q_1+k_1)+(q_4+q_1|q_{14}-k_1)\]
\beq
 -(-q_4+k_2|q_1+k_1+k_2)+(-q_{14}|-q_1-q_4-k_1)\Big].
\eeq

\[Y_4=
\frac{1}{4}(q_1-q_4+k_2|-q_4+k_1)
+\frac{1}{4}(-q_1+q_4-k_2|-q_1-k_1)\]
\beq
-\frac{1}{4}(-q_{14}|-q_1-k_1)
-\frac{1}{4}(q_{14}|q_4+k_1-k_2).
\eeq

\beq
Y_5=
-(q_4|q_4+k_1) -(-q_4+k_2|-q_4-k_1)+(k_2|-k_1).
\eeq

\[Y_6=
-\frac{1}{4}(q_4-q_1-k_2|q_4-k_1-k_2)
-\frac{1}{4}(-q_4+q_1+k_2|-q_4-k_1)\]
\[+\frac{1}{4}(q_{14}|q_4-k_1-k_2)
+\frac{1}{4}(-q_{14}|-q_4-k_1)\]
\beq
+\frac{1}{2}(-q_4+k_2|-q_4+k_1)-\frac{1}{2}(-q_4|-q_4+k_1)
-\frac{1}{2}(k_2|k_2-k_1).
\eeq
(We recall that $q_{14}=q_1+q_4$).

In their sum they contain  42 functions $F$. However more cancellations
occur in the sum $\sum_i(X_i-Y_i)$. Doing this sum we finally find
the total contribution to the BB cross-section from our diagrams as
\[
I^{BB}_R=V(q_1,q_4|q_1+k_2,q_4-k_2)\]\beq
 \Big(-2(q_1-q_4+k_2|k_1)+2(q_{14}|k_1)-4(q_1+k_2|k_1)+4(q_1|k_1)\Big).
\label{ibbr}
\eeq

It is instructive to see contributions of which diagrams survive in the
final expression (\ref{ibbr}). One finds that contributions $X_2$
from transitions
$2\to 4$ in the $D$ configuration Fig. \ref{fig8} and $Y_4$ from transitions
$4\to 4$ in the
single cut $DC$ configuration Fig. \ref{fig13} completely cancel.
From contributions
$X_5$, $Y_5$ and $Y_6$ only a single diagram (plus its conjugate)
gives contribution in each
case, namely Fig. \ref{fig14},$1+3$, Fig. \ref{fig15},$1+3$ and
Fig. \ref{fig17},$5$. On the other hand for transitions SD $2\to 4$ and
DC $3\to 4$ all diagrams in Fig. \ref{fig9} and
Fig. \ref{fig16} respectively  contribute to the final expression
(\ref{ibbr}).

Cross-section $I^{BB}_R$ is to be compared with the corresponding
contribution
from the KT cross-section. The latter is obtained from $I^{KT}$ by removing
one half of the first term in $I_1^{KT}$, since it is reproduced
by $I^{BB}_2$, and
changing sign of $I_3^{KT}$
due to contribution $I^{BB}_1$, Eq. (\ref{ibb1}).
So we have to compare $I^{BB}_R$ with
\[ \tilde{I}^{KT}=
-2(k_2|k_1)(q_{14}|q_1-k_2)+2(q_{14}|k_2)(q_1|k_1)
+2(q_{14}|k_2)(q_4|k_1)\]
\[
+4(q_1|-k_2)(q_{14}+k_2|k_1)+4(q_4|-k_2)(q_{14}+k_2|k_1)-
4(q_{14}|-k_2)(q_{14}+k_2|k_1)\]\beq
-2(q_1|-k_2)(q_{14}|k_1)
-2(q_4|-k_2)(q_{14}|k_1)+2(q_{14}|-k_2)(q_{14}|k_1),
\label {iktt}
\eeq
where for comparison we write all 9 terms separately enumerating them
as (KT1,...,KT9).

To compare the two expressions we present
\beq
V(q_1,q_4|q_1+k_2,q_4-k_2)=(q_{14}|q_1+k_2)-(q_1|-k_2)=(q_4|k_2)
\label{vv}
\eeq
and in the term in (\ref{ibbr}) which comes from the first term
in (\ref{vv}) shift the integration variable
$k_2\to k_2-q_1$. Then after dropping contributions which do not depend on
either $q_1$ or $q_4$ we find 10 terms
\[
I^{BB}_R=(q_{14}|k_2)\Big(-2(-q_4+k_2|k_1)+2(q_{14}|k_1)-4(k_2|k_1)+
4(q_1|k_1))\Big)
\]
\[
-(q_1|-k_1)\Big(-2(q_1-q_4+k_2|k_1)+2(q_{14}|k_1)\Big)\]\beq
-(q_4|k_2)
 \Big(-2(q_1-q_4+k_2|k_1)+2(q_{14}|k_1)-4(q_1+k_2|k_1)+4(q_1|k_1)\Big),
\label{ibbr1}
\eeq
which we enumerate as (BB1,...,BB10).

We observe immediately that (BB2)=(KT9), (BB4)=(KT2)+(KT3), (BB6)=(KT7),\\
(BB8)=(KT8). Changing in (BB5) and (BB7) $q_4\to -q_4$ we find
(KT4)+(KT5)=\\2(BB5)+2(BB7). Changing in (KT6) we find
(KT6)$=-4(q_{14}|k_2)(k_2|k_1)=$(BB3). Shifting in (KT1) $k_2\to k_2+q_1$
we have (KT1)=$-2(k_2+q_1|k_1)(q_{14}-k_2)$. Changing $k_2\to -k_2$ and
interchanging $q_1$ and $q_4$ we find (KT1)=(BB1).
At this point the difference between  $I_R^{BB}$ and $\tilde{I}^{KT}$
reduces to (BB9)+(BB10)$-$(1/2)[(KT4)+(KT5)].
Now we observe that  shift $k_2\to k_2-q_1$ transforms
\[ (q_1|-k_2)(q_{14}+k_2|k_1)\to (q_1|k_2)(q_4+k_2|k_1)\]
so that the interchange $q_1\leftrightarrow q_4$ gives
(1/2)[(KT4)+(KT5)]=(BB9). So the total difference between
$I_R^{BB}$ and $\tilde{I}^{KT}$ finally reduces to a single term
(B10):
\beq
 I_R^{BB}-\tilde{I}^{KT}=2(q_4|k_2)(q_1|k_1)+2(q_1|k_2)(q_4|k_1),
\label{final}
\eeq
where we have symmterized in $q_1$ and $q_4$. However one can prove
(see Appendix 2.) that after integration over $k_2$, $q_1$ and $q_4$
with the pomeron wave functions this term vanishes. So in the end
the inclusive cross-sections found in the BB and KT approaches
completely coincide.

\section{Conclusions}
We have performed direct calculations of the inclusive cross-section
for gluon production in both KT and BB approaches in the
NLO. Our result shows that these cross-sections are  identical.
This result may look astonishing. In fact all diagrams considered
for $I_R^{BB}$ in the BB approach involve in the $t$-channel
three or four intermediate reggeized gluons in the irreducible
colorless states: the BKP states. Presence of these states together
with the impossibility to use the so-called bootstrap relation due to
the fixed observed gluon momentum $k_1$ was the reason why in ~\cite{BSV}
a very complicated equation was derived for
the inclusive cross-section, which had several parts additional to the one
corresponding to the KT cross-section with explicit contribution
from the BKP states. Our results show that all these additional terms
cancel at least in the first nontrivial order in $\alpha_s\ln s$
(and of course in the limit $N\to\infty$). Strictly speaking this does
not exclude that the BKP contribution starts from the NNLO, although
we think it quite improbable. In any case the comparison of the
BB and KT cross-sections in the NNLO is certainly out of the question
due to the scale of the corresponding calculations. So our final
conclusion is that in spite of the presence of a lot of additional
BKP contributions the BB and KT cross-sections are identical.
This is very good news because in the opposite case one would have
to calculate the BKP Green functions for their evolution, which is
hardly feasible. In contrast, the KT cross-sections can be found
exclusively in terms of the standard BFKL pomeron and triple pomeron
vertex, which, as is well-known, even admits introduction of the running
coupling.

\section{Acknowledgements}
The author highly appreciaites discussions with J.Bartels, G.P.Vacca
and especially with Yu.Kovchegov, who initiated and have kept a constant
attention and interest in this work. This work was supported by
grants RFFI 09-012-01327-a and RFFI-CERN 08-02-91004.

\section{Appendix 1. The KT cross-section in terms of BB diagrams}
The BB diagrammatic representation of the KT cross-section is just
transformation to the momentum space.

The basic formula for the cross-section is ~\cite{bra2}
\beq
I(y,k,b)\equiv \frac{(2\pi)^3d\sigma}{dyd^2kd^2b}=
\int d^2r P(Y-y,r)V_k(r)\Big(2\Phi(y,r,b)-\Phi^2(y,r,b)\Big),
\label{defi}
\eeq
where
\beq
V_k(r)=\frac{4\alpha_s N}{k^2}\overleftarrow{\Delta}e^{ikr}\overrightarrow
{\Delta}.
\eeq
The LO contribution to the inclusive cross-section on two
centers comes from two sources: the quadratic term $\Phi^2$ and the
second term of the Glauber expansion of the initial target factor
\[ \Phi(y-0,r)=\Phi_0(r)-\frac{1}{2}\Phi_0^2(r).\]
So the LO contribution  is
\beq
I^{LO}_2=-2\int d^2rP_0(r)V_k(r)\Phi_0^2(r).
\eeq
where $P_0(r)$ is the lowest order contribution to the projectile
factor. Transformation to the momentum space gives
\beq
I^{LO}_2=-8\alpha_s N_c\int \frac{d^2q_1d^2q_4}{(2\pi)^4}
P_0(q_{14}+k)(q_{14}+k)^4(q_{14}|k)\Phi_0(q_1)\Phi_0(q_4).
\eeq
Putting the projectile and target factors to unity
\[P_0(q_{14}+k)(q_{14}+k)^4\to 1,\ \ \Phi_0(q)\to 1\]
and suppressing  integration over $q_1$ and $q_4$ with weight
$1/(2\pi)^2$ each we get
\beq
I^{LO}=-8\alpha_s N_c(q_{14}|k).
\eeq
Direct calculation of the correspondibg sum of BB diagrams with
$V$ and $W$ given by (\ref{defv}) and (\ref{defw}) correspondingly
gives
\beq
I^{BB,LO}=-4(q_{14}|k),
\label{ibblo}
\eeq
which fixes the relation between the sum of all BB diagrams $I^{BB}$
calculated with (\ref{defv}) and (\ref{defw}) and
the inclusive cross-section as
\beq
I(y,k,b)=2\alpha_s  N_cI^{BB}.
\eeq
(This confirms the relation established in a more general manner
in Section 3).

In the KT cross-section contribution (\ref{ibblo}) splits into
two equal parts, one of them corresponding to the second term in the
Glauber expansion of the initial condition and the other to the term
quadratic in $\Phi$. They evolve differently with the BFKL kernel, so that
in the NLO they become different and correspond to terms $I_2^{KT}$
and $I_3^{KT}$ given by (\ref{ikt2}) and (\ref{ikt3}) with the suppressed
factor and integrations given by (\ref{ink2}).

The contribution
coming from the first term in (\ref{defi}) linear in $ \Phi$ carries a
twice greater coefficient. For two scattering centers
it can be taken directly from the expression for
the for the forward triple pomeron amplitude ~\cite{bra2},
which in its turn uses the normalization introduced in ~\cite{bar}:
\[
T_{3\to 3}=-is\frac{1}{2\pi}\int dy\int\frac{d^2k_1}{(2\pi)^2}
\frac{d^2q_1}{(2\pi)^2}\frac{d^2q_4}{(2\pi)^2}\]\beq
\Gamma(q_1,-q_1,-q_4,q_4|k_1,-k_1)P(Y-y,k_1)P_1(y,q_1)P_4(y,q_4),
\eeq
where the splitting vertex is defined according to
\[
\Gamma(q_1,-q_1,-q_4,q_4|k_1,-k_1)=\tilde{g}^2
 k_1^4\Big(-\tilde{g}^2N_cW(q_1,-q_{14},q_4|k_1,-k_1)-\]
 \beq(2\pi)^3\delta^2(k_1-q_1)
\Big(\omega(-q_{14})-\omega(-q_1)\Big)-(2\pi)^3\delta(-k_1-q_4)
\omega(-q_4)\Big).
\label{gamma}
\eeq
Emission from the upper pomeron implies fixing integration over one of the
intermediate pomeron interaction inside, given by $-Vd^2k/(2\pi)^3$
So in the NLO we get Eq. (\ref{ikt1}), using (\ref{phipsi}),
suppressing the lower legs and integration over them and doubling
according to (\ref{defi}).
The overall sign in $I_1^{KT}$ takes into account signs of the insertion
of the gluon emisssion operator  ($-V$) and of the 3-Pomeron
vertex ~(see ~\cite{brava}).

\section{Appendix 2. A useful property of the BFKL pomeron}
In our notatiion he BFKL evolution equation reads
\beq
\pd_y P(y,q)=
\frac{g^2N}{8\pi^3}\int d^2\kappa \Big(2(q-\kappa|q)P(y,q-\kappa)
-(q|\kappa)P(y,q)\Big).
\label{eq2}
\eeq
We use the property that
the pomeron vanishes as the two reggeized gluons are located at the same
spatial point.
\beq
\int d^2q P(y,q)=0:
\label{id1}
\eeq
Integrating Eq. (\ref{eq2}) over $q$ we get
\beq
\int d^2q\int d^2\kappa \Big(2(q-\kappa|q)P(y,q-\kappa)
-(q|\kappa)P(y,q)\Big)=0.
\label{eq3}
\eeq
In the first term we shift the integration variable $q\to q+\kappa$
to obtain
\beq
\int d^2q\int d^2\kappa \Big(2(q|q+\kappa)
-(q|\kappa)\Big)P(y,q)=0.
\label{eq31}
\eeq
However $(q|q+\kappa)=(q|-\kappa)$ and changing in the first term
$\kappa\to -\kappa$ we find  the  identity
\beq
\int d^2q\int d^2\kappa (q|\kappa)P(y,q)=0.
\label{eq4}
\eeq
It means that apart from identity (\ref{id1})
one also has
\beq
\int d^2q \ln q^2 P(y,q)=0.
\label{id2}
\eeq

\end{document}